\documentclass[english,aps,prb]{revtex4}
\usepackage[latin1]{inputenc}
\usepackage{subfigure}
\usepackage{graphicx}
\usepackage{amssymb}

\makeatletter

\providecommand{\tabularnewline}{\\}

\usepackage{subfigure}
\usepackage{graphicx}

\usepackage{babel}
\makeatother
\begin{document}

\preprint{This line only printed with preprint option}

\title{Conduction electrons and the decoherence of impurity-bound electrons in a semiconductor}

\author{Kuljit S. Virk}

\email{kvirk@physics.utoronto.ca}

\affiliation{Department of Physics and Institute for Optical Sciences, University of Toronto, 60 St. George St., Toronto}

\author{J.E. Sipe}

\email{sipe@physics.utoronto.ca}

\affiliation{Department of Physics and Institute for Optical Sciences, University of Toronto, 60 St. George St.,
Toronto}

\begin{abstract}
We study the dynamics of impurity bound electrons interacting with
a bath of conduction band electrons in a semiconductor. Only the exchange
interaction is considered. We derive master equations for the density
matrices of single and two qubit systems under the usual Born and
Markov approximations. The bath mediated RKKY interaction in the two
qubit case arises naturally. It leads to an energy shift significant
only when the ratio ($R_{T}$) of the inter-qubit distance to the
thermal deBroglie wavelength of the bath electrons is small. This
bath mediated interaction also has a profound impact on the decoherence
times; the effect decreases monotonically with $R_{T}$. 
\end{abstract}
\maketitle

\section{introduction}

The dynamics of the impurity bound electrons in semiconductors have
been studied previously in the context of population relaxation \cite{Abrahams,Bardeen,Honig}.
Experimental studies of electron spin relaxation were central to understanding
various mechanisms of relaxation of nuclear and electron spins in
$^{31}$P atoms in Si. Since then, theoretical work has exhaustively
considered the effects of various interactions among electrons and
atoms on the relaxation of populations in different quantum states.
Various researchers have considered the effects of spin-orbit coupling,
hyperfine, exchange, Coulomb, and dipolar interactions. Current interest
in the unusually long relaxation times predicted and observed in these
systems stems from possible applications in solid-state quantum computing.
The seminal proposal by Kane \cite{kane} to use $^{31}$P atoms embedded
in Si crystal as qubits identifies one strategy for the realization
of quantum computer hardware in the solid state.

Recent investigations of decoherence mechanisms in these schemes assume
the presence of a strong magnetic field and low temperatures, typically,
a few mK; this prevents spontaneous spin flips by breaking the degeneracy
of bound electron spins \cite{kane,x_sarma,koiller_sarma,loss_vincenzo,rogerio_sarma}.
The dominant decoherence mechanisms in this case include the hyperfine
and dipolar interaction of electrons with nuclei, and dipolar-dipolar
interaction among the electrons. Recent calculations \cite{koiller_sarma,x_sarma}
suggest that at low temperatures decoherence is dominated by spin
diffusion induced by hyperfine and dipolar interactions of qubits
with the spin bath of Si atoms. In principle, and to a high degree
in practice, this mechanism can be eliminated by using ultra pure
$^{28}$Si crystals. But it would be desirable to operate solid state
quantum devices at low magnetic fields and higher temperatures, and
the price for this is the importance of additional decoherence channels
in the system. One of these is due to the interaction of qubits with
a bath of conduction electrons. Studies of decoherence due to this
channel have appeared in literature only recently \cite{flips,RKKY}.
The problem investigated by Kim \emph{et al.} \cite{flips} concerns
the exchange interaction of a qubit with a spin polarized one dimensional
stream of electrons. However, it only addresses the spin flip rates
of the qubit, and uses a bath that is physically different from an
unpolarized gas of conduction electrons typically found in a semiconductor.
The latter type of bath is used by Rikitake \emph{et al.} \cite{RKKY}
who study the effects of decoherence on the RKKY interaction of two
qubits in a bath consisting of a non-interacting degenerate electron
gas. 

In this paper we consider a simple model, similar to that of Rikitake
\emph{et al}. \cite{RKKY}, in which the spins of electrons bound
to donor atoms act as qubits and scatter the conduction electrons
via the exchange interaction. For reasonably low donor densities,
we show that the conduction electrons form a Boltzmann gas at all
temperatures. Therefore, in a Kane type model only a classical distribution
need be considered. The temperature is considered high enough that
the ratio of bound to free electron density ensures that interactions
among qubits are negligible compared to their exchange interaction
with conduction electrons. For P atoms in Si, these assumptions are
satisfied, in the absence of magnetic fields, for donor densities
of the order of $10^{16}cm^{-3}$ or lower. The nature of the bath has important consequences for the temperature
dependence of decoherence. Thus the results in the present paper are in
contrast to those in of Rikitake \emph{et al.}, despite similar master equations obtained
in both. Furthermore, the present paper addresses the Kane model more
concretely and makes stronger connection between the parameters of the
equations derived and fundamental properties of semiconductors.

In the following, we first present a full master equation for the
density matrix of a single qubit and obtain an intuitive analytical
result for the decoherence and relaxation times. The result is similar
in form to the phenomenological result obtained by Pines \emph{et
al}.\cite{Bardeen} for the relaxation rates under conditions similar
to those considered here. We then
derive the master equation for a system of two mutually non-interacting qubits, and study the decoherence
due to their interaction with the conduction electron bath. We plan
to include the effects of interactions among qubits in a future paper.

\section{\label{sec:Model-and-Formal}Model and Formal Equations}

We consider a silicon lattice at nonzero temperatures doped with a
density $n_{D}$ of phosphorus atoms. Each P atom donates an electron,
which either becomes a conduction electron or is captured by another
ionized P atom forming an {}``atom'' with hydrogen-like properties.
The captured electrons are usually in s-states with a {}``Bohr radius''
of about 25 angstroms and a binding energy of about 0.044 eV \cite{rogerio_sarma,kohn}.
The conduction electrons form a gas of approximately free particles
with an effective mass of $0.2m_{e}$, where $m_{e}$ is the bare
mass of the electron. The density of the gas builds up (from zero
at $T=0$) as temperature rises and more donors are ionized. A simple
statistical mechanics analysis shows that the expected number density
of bound electrons at temperature $T$ is $n_{b}=n_{D}[1+\frac{1}{2}z^{-1}\exp(-E_{b}/k_{B}T)]^{-1}$,
where $z$ is fugacity of the total system comprised of bound and
unbound electrons, $E_{b}$ is the energy needed to excite a bound
electron into the conduction band, and $k_{B}$ is Boltzmann's constant.
The conduction electron gas has a number density of\begin{eqnarray}
n_{c}(T) & = & 2\lambda_{T}^{-3}F_{3/2}(z),\label{eq:nc}\end{eqnarray}
 where $F_{3/2}(z)=\int_{0}^{\infty}dx\sqrt{x}(z^{-1}e^{x}+1)^{-1}$
is the Fermi-Dirac function, and $\lambda_{T}=\hbar(2\pi/mk_{B}T)^{1/2}$
is the thermal deBroglie wavelength. Forcing $n_{c}+n_{b}=n_{D}$,
we find\begin{eqnarray}
F_{3/2}(z)[2z+e^{-E_{b}/k_{B}T}] & = & \frac{1}{2}n_{D}\lambda_{T}^{3}e^{-E_{b}/k_{B}T}.\label{eq:fugacity}\end{eqnarray}
 For the parameters chosen, the product $\lambda_{T}^{3}e^{-E_{b}/k_{B}T}<10^{-24}cm^{3}$
for temperatures below 300 K, which implies that for $n_{D}\approx10^{16}cm^{-3}$,
we can take $z\ll1$ and $F_{3/2}(z)\approx z$ . Consequently, the
distribution of conduction electron gas remains Boltzmann down to
$T=0$.

We simplify the picture by assuming the qubits to be in s-states,
with only the spin acting as their degree of freedom, and suppose
there are no external fields breaking the degeneracy of the spin states.
The conduction electrons collide occasionally with the qubits elastically;
we assume they do not excite them into higher states. However, the
spins of the two may become entangled or even exchanged. Since the
conduction electron moves throughout the crystal interacting with
many electrons and atoms before the qubit scatters another conduction
electron, it loses its coherence much faster than the qubits, and
may be thought to be in an incoherent superposition of momentum eigenstates
over the timescale of interest. Thus it is also independent of other
bath electrons, while being on the same footing as them. This allows
us to study only the case of interaction between a qubit and one mixed
state conduction electron, and multiply by the number of the latter
in the end result. Thus we can take our Hamiltonian as

\begin{eqnarray}
H & = & \frac{\mathbf{p}^{2}}{2m}+V,\label{eq: sec2 H}\end{eqnarray}
 where \textbf{p} is the momentum of the conduction electron and $m$
is its effective mass. The operator $V$ is the interaction Hamiltonian
acting on the electron and the qubit(s). We are concerned only with
the exchange interaction, and study two cases. We first consider a
single qubit at the origin for which the interaction takes the form
\begin{eqnarray}
V & = & Jr_{0}^{3}\delta(\mathbf{r})\mathbf{S}\cdot\mathbf{s},\label{eq:V 1}\end{eqnarray}
where $J$ is the exchange coefficient and $r_{0}$ is an effective
Bohr radius characterizing the size of the qubit. The spin operators
\textbf{$\mathbf{S}$} and \textbf{s} act on the qubit and the conduction
electrons respectively. We also consider two mutually non-interacting
qubits for which the interaction is \begin{eqnarray}
V & = & Jr_{0}^{3}\left[\delta(\mathbf{r}-\frac{1}{2}\mathbf{R})\mathbf{S}_{1}\cdot\mathbf{s}+\delta(\mathbf{r}+\frac{1}{2}\mathbf{R})\mathbf{S}_{2}\cdot\mathbf{s}\right],\label{eq:V 2}\end{eqnarray}
where $\mathbf{S}_{1}$ and $\mathbf{S}_{2}$ are the qubit spin operators,
and the two qubits are placed symmetrically at $\pm\mathbf{R}/2$.
To treat more than two qubits, similar terms would be added to $V$,
with delta functions centered at the respective qubit locations. Despite
the absence of direct exchange interaction between them, the two qubits
can still exchange spins \emph{via} indirect exchange interaction.
Physically this occurs when the bath remains correlated long enough
for the conduction electron to link two qubits; the indirect exchange
coupling between qubits can arise even when they are too far away
to have significant direct exchange interaction. This coupling is
significant only when the inter-qubit distance is much shorter than
the coherence length of the bath, which is approximately the thermal
deBroglie wavelength $\lambda_{T}$. The results found here depend
naturally on these two important length scales.

In addition, there are also direct interactions between the qubits.
The important qubit-qubit interactions involve the exchange interaction
between bound electrons, the hyperfine, dipolar, and spin-orbit coupling
of these electrons to the bath of Si nuclei. For low doping densities
the first of these can be ignored as a starting point. For a donor
density of $n_{D}=10^{16}cm^{-3}$, and $T=100K$, the inter-donor
distance is $R\approx50$ nm, whereas the mean radius of the electron
orbits is $r_{0}\approx 2.5$ nm. Thus the exchange energy is small,
as we expect little overlap between the qubit wave functions. The
second and third have been studied by various authors in the context
of both relaxation and decoherence rates \cite{Abrahams,Bardeen,Feher,Honig,Tyrishkin,koiller_sarma,x_sarma}.
The hyperfine and dipolar terms can be made arbitrarily small by purifying
the Si samples to contain only the $^{28}$Si isotope. For natural
Si, which contains 95.33\% $^{28}$Si, it is estimated that the hyperfine
and dipolar couplings combined are of the order of $10^{-7}$ eV or
less \cite{rogerio_sarma}. This is minute compared to the exchange
interaction on the order of meV that we consider. Spin-orbit coupling
is likewise small, and we neglect these additional effects in this
preliminary investigation.

We point out that the model has important differences to the  one used by Chang \emph{et al.} in their general study of dissipative dynamics of a two-state system\cite{chang}. They consider
a biased qubit that is coupled to the bath via $S_z$ only, and the coupling
induces no spin-flips in either the bath or the qubit. Furthermore, spin flips are introduced
by a tunneling parameter that is independent of the bath state. This is clearly not the case
in (\ref{eq:V 1}-\ref{eq:V 2}) where the isotropic coupling of system and bath causes joint spin flips
in the two subsystems. A more general case of Brownian motion in a fermionic environment has also been studied in several papers by Chen\cite{ychen_fermibrownian,ychen_adiabatic,ychen_surface}, who mainly focused on mapping between fermionic and bosonic environments. None of these studies discusses decoherence directly, and it is highly nontrivial to extend the results of these papers to arrive at those in this work.

We first develop a general equation for the density matrix $\rho(t)$
of a system coupled to a bath and the full system evolving via the
Hamiltonian in (\ref{eq: sec2 H}). The following notation is used.
We label the conduction electron states by $\left|\alpha\mathbf{p}\right\rangle =\left|\alpha\right\rangle \otimes\left|\mathbf{p}\right\rangle $,
where $\left|\alpha\right\rangle $ and $\left|\mathbf{p}\right\rangle $
are the eigenstates of $s^{z}$ and the momentum operator of the conduction
electron respectively. The kinetic energy of the bath electron is
denoted by $\hbar\omega_{p}=p^{2}/2m$, and the normalized system
states are labelled using Roman letters. For a general interaction
$V$, we write the interaction Hamiltonian in the interaction picture
as\begin{eqnarray}
H_{I}(t) & = & \sum_{ij}\left|i\right\rangle \left\langle j\right|\otimes\sum_{\alpha,\alpha'}\sum_{\mathbf{p},\mathbf{p}'}e^{i(\omega_{p}-\omega_{p'})t}\left|\alpha\mathbf{p}\right\rangle \left\langle \alpha\mathbf{p}\right|V_{ij}\left|\alpha'\mathbf{p}'\right\rangle \left\langle \alpha'\mathbf{p}'\right|\label{eq:sec2 Hint}\end{eqnarray}
 Here we defined bath operators $V_{ij}=\left\langle i\right|V\left|j\right\rangle $
which depend parametrically on system states. It is shown in the Appendix
that for an unpolarized bath under the Born approximation, the density
matrix $\rho$ of the system is given by\begin{eqnarray}
\dot{\rho} & = & -i\omega_{RKKY}\sum_{ijkl}\Gamma_{ijkl}[\eta_{ij}\eta_{kl},\rho]+\gamma\sum_{ijkl}\Lambda_{ijkl}(2\eta_{kl}\rho\eta_{ij}-\eta_{ij}\eta_{kl}\rho-\rho\eta_{ij}\eta_{kl}),\label{eq: sec2 rho1}\end{eqnarray}
 where the operators $\eta_{ij}=\left|i\right\rangle \left\langle j\right|$
are system operators. The tensors $\Gamma$ and $\Lambda$, and the
constants $\gamma$ and $\omega_{RKKY}$, result from specializing
(\ref{eq:app M final}) and (\ref{eq:app N final}) to the two the
forms of $V$ shown in (\ref{eq:V 1}) and (\ref{eq:V 2}). It is
evident that $\Gamma$ is involved with a unitary evolution of $\rho(t)$
and $\Lambda$ with the decoherence processes. Thus we label the first
sum as the \emph{unitary term} and the second as the \emph{dissipative}
or \emph{non-unitary}. The frequency $\omega_{RKKY}$ is a shift resulting
from the RKKY interaction between the two qubits. No such term arises
in the single qubit problem. The constant $\gamma^{-1}$ is an interaction
time constant of the system and the bath, (see appendix for derivation);
$\gamma$ and $\omega_{RKKY}$ are defined as

\begin{eqnarray}
\gamma & = & 4(2\pi)^{-\frac{3}{2}}n_{c}(T)(Jr_{0}^{3}m\hbar^{-2})^{2}\sqrt{\frac{k_{B}T}{m}},\label{eq:sec2 gamma}\\
\omega_{RKKY}(\mathbf{R}) & = & -(2\pi)^{3}\hbar^{-1}(Jr_{0}^{3})^{2}\chi(\mathbf{R})\label{eq:sec2 omega rkky}\\
 & = & \frac{4}{\pi}n_{c}(T)(Jr_{0}^{3}m\hbar^{-2})^{2}\frac{\hbar}{mR}e^{-\pi(2R/\lambda_{T})^{2}}.\label{eq:sec2 omega boltz}\end{eqnarray}
 In the definition (\ref{eq:sec2 omega rkky}) of $\omega_{RKKY}$,
$\chi(\mathbf{r})$ is the free electron susceptibility, and the expression
in (\ref{eq:sec2 omega boltz}) specializes this definition to a Boltzmann
distribution for the conduction electron gas. Apart from dimensionless
factors, $\gamma$ is a product of the thermal flux of conduction
electrons and an {}``effective area'' $(Jr_{0}^{3}m\hbar^{-2})^{2}$
over which the bound electron experiences this flux. On the other
hand, the thermal flux in the definition of $\omega_{RKKY}$ is replaced
by $n_{c}\hbar(mR)^{-1}\exp(-4\pi R_{T}^{2})$, which depends on the
inter-qubit distance and the thermal deBroglie wavelength of bath
electrons. Both these rates are given by a mean number of scattering
events, where the scattering occurs \emph{via} the exchange interaction.
The $1/R$ divergence of $\omega_{RKKY}$ is typical of the $\chi(\mathbf{r})$
in three dimensions \cite{larsen,kim95} but the spatial damping factor
in the case of Boltzmann distribution is Gaussian as opposed to exponential
in the degenerate limit \cite{kim95}. Furthermore, the increase in
decoherence with $n_{c}$ and $J$ is accompanied by a correspondingly
faster unitary evolution, as the ratio $\gamma^{-1}\omega_{RKKY}$
is independent of both these parameters.

\section{Application to two specific cases}

\subsection{Dynamics of a single qubit\label{sub:Single-Qubit}}

This is the simplest application of the general expressions given
above, and we use it to estimate the temperature dependence of relaxation
and decoherence times. The tensors $\Gamma$ and $\Lambda$ in this
case are equal and their tensor elements are found to be\begin{eqnarray}
\Lambda_{ijkl} & = & \frac{1}{2}\left\langle i\right|S^{z}\left|j\right\rangle \left\langle k\right|S^{z}\left|l\right\rangle +\frac{1}{4}\left\langle i\right|S^{+}\left|j\right\rangle \left\langle k\right|S^{-}\left|l\right\rangle +\frac{1}{4}\left\langle i\right|S^{-}\left|j\right\rangle \left\langle k\right|S^{+}\left|l\right\rangle ,\label{eq:lambda}\end{eqnarray}
 where $S^{\pm}=S^{x}\pm iS^{y}$. Substitution of this expression
in (\ref{eq: sec2 rho1}) yields the master equation for a single
qubit explicitly in the Lindblad form. The unitary term becomes $[S^{2},\rho]$,
and it vanishes because $S^{2}$ is proportional to identity. We define
a superoperator $\mathcal{L}(.)$ such that $\mathcal{L}(A)\rho=2A\rho A^{\dagger}-A^{\dagger}A\rho-\rho A^{\dagger}A$,
where $A$ is a general operator, and obtain\begin{eqnarray}
\dot{\rho} & = & \frac{\gamma}{2}[\mathcal{L}(S^{x})\rho+\mathcal{L}(S^{y})\rho+\mathcal{L}(S^{z})\rho]\label{eq:1qbit lind}\\
 & = & -\frac{\gamma}{2}([S^{x},[S^{x},\rho]]+[S^{y},[S^{y},\rho]]+[S^{z},[S^{z},\rho]]).\label{eq:1qbit double comm}\end{eqnarray}
 The latter equality follows due to the Hermiticity of the spin operators.
Besides ensuring complete positivity and trace preservation, this
form is also invariant under arbitrary rotations of the spin. Therefore
we may consider any pure initial state to be a spin up state in the
$S^{z}$ basis of a suitable coordinate frame. In the Bloch formalism,
these states reside on the surface of a unit sphere and their loss
of purity is described by the decay in the length of the vector representing
them. From (\ref{eq:1qbit double comm}) we then find that the initial
decay of purity is equal to $d\mathsf{Tr}[\rho^{2}]/dt=-\gamma$ for
all pure states.

To study the evolution of a general state, we study the Bloch vector
with components $u=\rho_{01}+\rho_{10}$, $v=i(\rho_{01}-\rho_{10})$,
and $w=\rho_{11}-\rho_{00}$. Here 0 identifies the {}``spin down''
state and 1 the {}``spin up'' state. From the master equation (\ref{eq:1qbit double comm})
it follows that\begin{eqnarray}
\dot{u} & = & -\gamma u,\,\,\,\dot{v}=-\gamma v,\,\,\,\dot{w}=-\gamma w.\label{eq:bloch}\end{eqnarray}

From these equations we find that the longitudinal and transverse
rates are $T_{1}^{-1}=T_{2}^{-1}=\gamma$. Thus the relaxation rate
is proportional to the collision rate of a qubit with a thermalized
bath of particles. The two rates are equal because the system is unbiased,
and both relaxation and decoherence arise only through elastic scattering.
The conventional relation $2T_{2}\leq T_{1}$ holds true only when
several distinct scattering processes determine $T_{2}$ while only
a subset of them is responsible for $T_{1}$. When no such distinction
exists we can have $T_{1}=T_{2}$ , as Bloch noted many years ago
\cite{Bloch}.

\begin{figure}

\caption{\label{fig:nc}Ratio of conduction electron density to donor density
as a function of temperature}

\includegraphics[%
  scale=0.4]{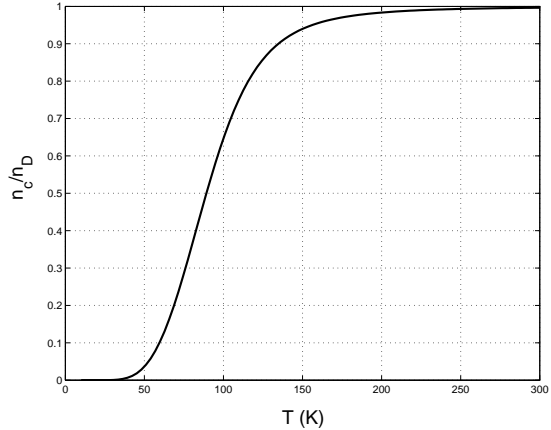}
\end{figure}

\begin{figure}

\caption{\label{fig:decoherence}Decoherence time for a single qubit as a
function of temperature (logarithm base 10).}

\includegraphics[%
  scale=0.4]{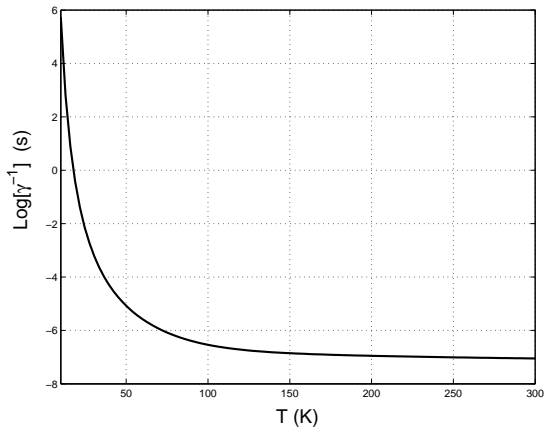}
\end{figure}

In Si:P with physical parameters defined in Section \ref{sec:Model-and-Formal},
we estimate $J\approx6$meV. Taking into account the temperature dependence
of $n_{c}(T)$, shown in Figure (\ref{fig:nc}), we plot $\log(\gamma^{-1})$
as a function of $T$ in Figure (\ref{fig:decoherence}) (the logarithm base is 10). It is evident
from the figure that above about 70 K, $\gamma^{-1}$ is less than
a microsecond, which means that conduction electrons present a significant
decoherence mechanism in this regime. The effectiveness of this channel
vanishes significantly at low temperatures due to the loss of conduction
electrons (as discussed in Section \ref{sec:Model-and-Formal}); $n_{c}$
is less than 1 percent of $n_{D}$ for $T<35$K. The $\sqrt{T}$ dependence
of the thermal flux, which is associated with a decrease in conduction
electron velocity with lowering temperature, also contributes to the
rapid decrease in $\gamma$ as temperature is decreased.

\subsection{Dynamics of two mutually non-interacting qubits\label{sub:Two-Qubits}}

\begin{figure}

\caption{\label{fig:xi}$\xi(R_T)$ as a function of $R_T=R/\lambda_T$}

\includegraphics[%
  scale=0.4]{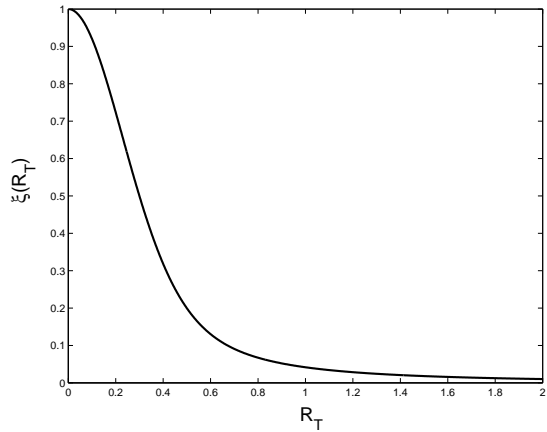}
\end{figure}

We now consider a system of two mutually non-interacting qubits that
relax and decohere \emph{via} the exchange interaction with the conduction
electrons. Again, the general dynamical map (\ref{eq: sec2 rho1})
describes the evolution upon specializing the tensors $\Lambda$ and
$\Gamma$ to (\ref{eq:V 2}) using the defining equations (\ref{eq:app M final})
and (\ref{eq:app N final}). As expected a contribution of the form
(\ref{eq:lambda}) for a single qubit comes from each member of the
system. We call this $\Lambda'$ and write it as

\begin{eqnarray}
\Lambda'_{ijkl} & = & \sum_{q=1}^{2}\frac{1}{2}\left\langle i\right|S_{q}^{z}\left|j\right\rangle \left\langle k\right|S_{q}^{z}\left|l\right\rangle +\frac{1}{4}\left\langle i\right|S_{q}^{+}\left|j\right\rangle \left\langle k\right|S_{q}^{-}\left|l\right\rangle +\frac{1}{4}\left\langle i\right|S_{q}^{-}\left|j\right\rangle \left\langle k\right|S_{q}^{+}\left|l\right\rangle ,\label{eq:lambda prime}\end{eqnarray}
 where $q=1,2$ labels each qubit. However, there also exist {}``cross-coupling''
terms which represent the process by which a conduction electron mediates
a spin exchange between the two qubits. These are given by $\Lambda''$
where \begin{eqnarray}
\Lambda''_{ijkl} & = & \frac{1}{2}\left\langle i\right|S_{1}^{z}\left|j\right\rangle \left\langle k\right|S_{2}^{z}\left|l\right\rangle +\frac{1}{4}\left\langle i\right|S_{1}^{+}\left|j\right\rangle \left\langle k\right|S_{2}^{-}\left|l\right\rangle +\frac{1}{4}\left\langle i\right|S_{1}^{-}\left|j\right\rangle \left\langle k\right|S_{2}^{+}\left|l\right\rangle \nonumber \\
 &  & +\frac{1}{2}\left\langle i\right|S_{2}^{z}\left|j\right\rangle \left\langle k\right|S_{1}^{z}\left|l\right\rangle +\frac{1}{4}\left\langle i\right|S_{2}^{+}\left|j\right\rangle \left\langle k\right|S_{1}^{-}\left|l\right\rangle +\frac{1}{4}\left\langle i\right|S_{2}^{-}\left|j\right\rangle \left\langle k\right|S_{1}^{+}\left|l\right\rangle .\label{eq:lambda prime2}\end{eqnarray}
 The two tensors add to form the tensors that appear in (\ref{eq: sec2 rho1}):
\begin{eqnarray}
\Lambda_{ijkl} & = & \Lambda'_{ijkl}+\xi(R_{T})\Lambda''_{ijkl},\label{eq:lambda q2}\\
\Gamma_{ijkl} & = & \Lambda''_{ijkl},\label{eq:gamma 2}\\
\xi(R_{T}) & = & \int_{0}^{\infty}xe^{-x}\mathrm{sinc}^{2}[R_{T}\sqrt{4\pi x}]dx = \frac{-i\mathrm{Erf}(iR_T\sqrt{4\pi})}{4R_T}e^{-4\pi R^2_T}  \label{eq:Irt}\end{eqnarray}
 where $R_{T}=R/\lambda_{T}$ is the inter-qubit distance measured
in units of the thermal deBroglie wavelength, and $\mathrm{Erf}(.)$ denotes the error function. The tensor $\Gamma$
depends only on $\Lambda''$ because the term corresponding to $\Lambda'$
commutes with $\rho(t)$, as can be verified from (\ref{eq: sec2 rho1}).
The dimensionless integral $\xi(R_{T})$ represents the strength of
indirect exchange coupling between the two qubits relative to their
direct exchange interaction with the conduction electrons. The most
dominant contribution to the integral comes from $x<\pi/4R_{T}^{2}$.
Thus we find, as expected, that the strength of indirect exchange
decreases rapidly as qubits move out of the coherence region of the
conduction electron, and the plot of $\xi(R_T)$ in Figure (\ref{fig:xi}) 
shows a monotonic decrease as $R_T$ increases. We point out that if the Fermi distribution were
applicable, $\xi(R,T)$ would be given by the same formula as above,
but with $xe^{-x}$ replaced by the $x(z^{-1}e^{x}+1)^{-1}$, where
$z$ is the fugacity of the gas. At vanishingly small temperatures,
the Fermi wavelength would then take the role of the thermal deBroglie
wavelength in setting the length-scale of indirect exchange. But we
stress that for the model presented in section \ref{sec:Model-and-Formal}
the Boltzmann distribution is relevant for all temperatures.

Complete positivity and trace preservation is guaranteed in the two
qubit case as well. Substitution of (\ref{eq:lambda prime}-\ref{eq:lambda prime2})
in (\ref{eq: sec2 rho1}) yields\begin{eqnarray}
\dot{\rho} & = & -\frac{i}{\hbar}[H_{eff},\rho]+\sum_{k,l=1}^{6}c_{kl}[2F_{k}\rho F_{l}^{\dagger}-\{ F_{l}^{\dagger}F_{k},\rho\}],\label{eq:Lidblad q2}\end{eqnarray}
 where we have defined an effective Hamiltonian \begin{eqnarray}
H_{eff} & = & \hbar\omega_{RKKY}\mathbf{S}_{1}\cdot\mathbf{S}_{2},\label{eq:Heff}\end{eqnarray}
 which causes unitary evolution due to system bath interaction. The
dissipative part contains six operators $(F_{1},F_{2},F_{3},F_{4},F_{5},F_{6})=(S_{1}^{z},S_{1}^{+},S_{1}^{-},S_{2}^{z},S_{2}^{+},S_{2}^{-})$,
and a $6\times6$ symmetric coefficient matrix \begin{eqnarray}
c & = & \,\,\left[\begin{array}{cc}
a & \xi a\\
\xi a & a\end{array}\right],\label{eq:c matrix}\\
a & = & \frac{1}{4}\left[\begin{array}{ccc}
2 & 0 & 0\\
0 & 1 & 0\\
0 & 0 & 1\end{array}\right].\label{eq:a matrix}\end{eqnarray}
 The matrix $c$ is positive semidefinite, as can be verified from
its non-negative eigenvalues for $0\leq\xi\leq1$. This is sufficient
to ensure complete positivity of (\ref{eq:Lidblad q2}). The equation
is reduced to Lindblad form by diagonalizing $c$ and obtaining an
orthonormal set of eigenvectors. Thus\begin{eqnarray}
\dot{\rho} & = & -\frac{i}{\hbar}[H_{eff},\rho]\nonumber \\
 &  & +\frac{1}{4}\gamma(1+\xi)[\mathcal{L}(S_{2}^{x}+S_{1}^{x})+\mathcal{L}(S_{2}^{y}+S_{1}^{y})+\mathcal{L}(S_{2}^{z}+S_{1}^{z})]\rho\nonumber \\
 &  & +\frac{1}{4}\gamma(1-\xi)[\mathcal{L}(S_{2}^{x}-S_{1}^{x})+\mathcal{L}(S_{2}^{y}-S_{1}^{y})+\mathcal{L}(S_{2}^{z}-S_{1}^{z})]\rho.\label{eq: 2qbit lindblad}\end{eqnarray}

An equation of the same form has been derived by Rikitake \emph{et al.} but with 
the parameters specific to a degenerate bath\cite{RKKY}. Let us now compare this map with the single qubit map (\ref{eq:1qbit lind})
by ignoring the unitary term. For the case $\xi=0$, $c$ is already
diagonal, and the spin operators of each qubit form the set of Lindblad
operators in the dynamical map. The map then consists of a sum of
maps (\ref{eq:1qbit lind}) for each qubit, which implies that the
reduced dynamics of each qubit is independent of the other. Physically,
at $\xi=0$, the coherence length of the bath is much smaller than
the inter-qubit distance, and therefore the qubits scatter bath electrons
independently of each other.

In the presence of indirect coupling, the Lindblad operators are not
the spin operators of the qubits but their sum $\mathbf{S}=\mathbf{S}_{2}+\mathbf{S}_{1}$,
and difference $\mathbf{\Delta}=\mathbf{S}_{2}-\mathbf{S}_{1}$. For
significant values of $\xi$, the coherence length of the bath covers
the inter-qubit distance. When scattering the bath electrons, it is
then reasonable to expect that the two qubits behave as a single entity
with total spin $\mathbf{S}$. In fact, in the extreme limit of full
coherence over the region containing the qubits, $\xi=1$, and the
map has exactly the same form as that of the single qubit map. The
difference operator, $\mathbf{\Delta}$, accounts for the deviation
from perfect coherence of conduction electrons at the two sites, and
allows independent evolution to take place; the magnitude of this
effect is of order $1-\xi$. This suggests that the singlet state
with a total spin zero should be free of dissipative dynamics whenever
$\xi=1$, and this is found to be the case in the solution of (\ref{eq: 2qbit lindblad})
presented below. Note that this is true not just within the context
of the Born approximation to the scattering amplitudes. Since $\xi=1$
strictly only for $\mathbf{R}=0$ (it is meaningless to consider $T=0$
as $n_{c}(0)=0$), the interaction becomes $V=J\delta(\mathbf{r})\mathbf{s}\cdot\mathbf{S}$.
The operator $\mathbf{S}$ has a nullspace with the singlet as its
only member, and therefore, the singlet stops interacting with the
bath when $\xi=1$, and remains pure indefinitely. More generally,
the purity is clearly long lived whenever $\lambda_{T}\gg R$, and
$\xi\approx1$. Note also that no state other than the singlet can
be in isolation because a nonzero total spin would always interact
with the bath electrons. Finally, the dynamics of the singlet was briefly 
considered by Rikitake \emph{et al.}. They find similar behavior for the singlet,
but where the thermal deBroglie wavelength is replaced by Fermi wavelength, as was
pointed out earlier for the case of a degenerate gas.

Let us now consider in detail a realistic case of intermediate values
of $\xi$. In order to proceed, we analytically solve (\ref{eq: 2qbit lindblad}),
which is done most conveniently in the Bell-basis representation of
$\rho(t)$, where the basis vectors are enumerated in the order they
are shown below: \[
\left\{ \frac{\left|00\right\rangle +\left|11\right\rangle }{\sqrt{2}},\,\,\frac{\left|00\right\rangle -\left|11\right\rangle }{\sqrt{2}},\,\,\frac{\left|01\right\rangle +\left|10\right\rangle }{\sqrt{2}},\,\,\frac{\left|01\right\rangle -\left|10\right\rangle }{\sqrt{2}}\right\} .\]
 The first (second) symbol corresponds to the first (second) qubit.
In this basis the operators $\mathbf{S}$ and $\mathbf{\Delta}$ take
a particularly simple form as shown in Table (\ref{table1}). It is
evident from the table and (\ref{eq: 2qbit lindblad}) that the diagonal
terms of the density matrix do not couple to the off-diagonal ones.
This simplifies the calculation and yields the following population
equations:\begin{eqnarray}
\frac{d}{dt}(\rho_{22}-\rho_{11}) & = & -\gamma(2+\xi)(\rho_{22}-\rho_{11}),\label{eq:sec2 diff1}\\
\frac{d}{dt}(\rho_{33}-\rho_{11}) & = & -\gamma(2+\xi)(\rho_{33}-\rho_{11}),\\
\frac{d}{dt}\sum_{i=1,2,3}\rho_{ii} & = & -2\gamma(1-\xi)\sum_{i=1,2,3}\rho_{ii}+\frac{3\gamma}{2}(1-\xi)\\
\frac{d}{dt}\rho_{44} & = & -2\gamma(1-\xi)\rho_{44}+\frac{\gamma}{2}(1-\xi).\label{eq:sec2 rho44}\end{eqnarray}
 The first two equations show that the populations within the triplet
manifold equilibrate to a uniform distribution at a rate of $\gamma(2+\xi)$.
The last two show that the population transfer between this and the
singlet manifold occurs at a rate of $2\gamma(1-\xi)$, confirming
our observation that the singlet ceases to evolve at $\xi=1$. The
off-diagonal elements couple only to their conjugates. Within the
triplet manifold they obey \begin{eqnarray}
\rho_{12}(t) & = & \left(\Re[\rho_{12}(0)]+i\Im[\rho_{12}(0)]e^{-\gamma(1+\xi)t}\right)e^{-\gamma t},\label{eq:sec2 rho12}\\
\rho_{13}(t) & = & \left(\Re[\rho_{13}(0)]+i\Im[\rho_{13}(0)]e^{-\gamma(1+\xi)t}\right)e^{-\gamma t},\\
\rho_{23}(t) & = & \left(\Re[\rho_{23}(0)]e^{-\gamma(1+\xi)t}+i\Im[\rho_{23}(0)]\right)e^{-\gamma t},\end{eqnarray}
 while the elements between the triplet and the singlet manifolds
obey
\begin{eqnarray}
\rho_{14}(t) & = &\Re[\rho_{14}(0)]\left\{ \cos(\omega't)-(i\omega+\gamma')\frac{\sin(\omega't)}{\omega'}\right\}e^{-\gamma(3-\xi)t/2}\nonumber\\
             &  &+i\Im[\rho_{14}(0)]\left\{ \cos(\omega't)-(i\omega-\gamma')\frac{\sin(\omega't)}{\omega'}\right\}e^{-\gamma(3-\xi)t/2},\\
\rho_{24}(t) & = &\Re[\rho_{24}(0)]\left\{ \cos(\omega't)-(i\omega-\gamma')\frac{\sin(\omega't)}{\omega'}\right\}e^{-\gamma(3-\xi)t/2}\nonumber\\
             &  &+i\Im[\rho_{24}(0)]\left\{ \cos(\omega't)-(i\omega+\gamma')\frac{\sin(\omega't)}{\omega'}\right\}e^{-\gamma(3-\xi)t/2},\\
\rho_{34}(t) & = &\Re[\rho_{34}(0)]\left\{ \cos(\omega't)-(i\omega-\gamma')\frac{\sin(\omega't)}{\omega'}\right\}e^{-\gamma(3-\xi)t/2}\nonumber\\
             &  &+i\Im[\rho_{34}(0)]\left\{ \cos(\omega't)-(i\omega+\gamma')\frac{\sin(\omega't)}{\omega'}\right\}e^{-\gamma(3-\xi)t/2}.
\end{eqnarray}
 Here we defined the difference of eigenvalues of $H_{eff}$ in the
triplet and singlet manifolds as $\hbar\omega=\left\langle k\right|H_{eff}\left|k\right\rangle -\left\langle 4\right|H_{eff}\left|4\right\rangle $
, where $k=1,2$ or $3$ labels the triplet states. The renormalized
frequency $\omega'=\sqrt{\omega^{2}-\gamma'^{2}}$, where $\gamma'=\gamma(1-\xi)/2$,
represents the oscillations caused by the unitary evolution resulting
from the RKKY interaction. It follows from Table (\ref{table1}) that
these oscillations are absent within the triplet manifold; they also disappear for $\omega<\gamma'$ in the above equations.
 We note that the off-diagonal elements
always decay at least with a rate of $\gamma$, and each of these
elements is uncoupled from all others. Hence dephasing between any
pair of Bell states proceeds independently of the rest of the states.

\begin{table}[b]

\caption{\label{table1}Bell basis representation of the spin operators $\mathbf{S}$
and $\mathbf{\Delta}$.}

\begin{tabular}{|c||c|c|c|}
\hline 
$j\rightarrow$&
$x$&
$y$&
$z$\tabularnewline
\hline
\hline 
$S^{j}$&
$\left|1\right\rangle \left\langle 3\right|+\left|3\right\rangle \left\langle 1\right|$&
$-i\left|2\right\rangle \left\langle 3\right|+i\left|3\right\rangle \left\langle 2\right|$&
$\left|1\right\rangle \left\langle 2\right|+\left|2\right\rangle \left\langle 1\right|$\tabularnewline
\hline 
$\Delta^{j}$&
$-\left|2\right\rangle \left\langle 4\right|-\left|4\right\rangle \left\langle 2\right|$&
$-i\left|1\right\rangle \left\langle 4\right|+i\left|4\right\rangle \left\langle 1\right|$&
$\left|3\right\rangle \left\langle 4\right|+\left|4\right\rangle \left\langle 3\right|$\tabularnewline
\hline
\end{tabular}
\end{table}

Equations (\ref{eq:sec2 diff1}-\ref{eq:sec2 rho44}) show that, for
$\xi\neq1$, the final state of the density matrix is the maximum
entropy state $\rho=\frac{1}{4}\mathbf{1}$. However, for $\xi=1$,
the relaxation between the singlet state and the triplet manifold
ceases, and the final state becomes $\rho_{44}^{final}=\rho_{44}(0)$
and $\rho_{ii}^{final}=\frac{1}{3}(1-\rho_{44}(0))$ for $i=1,2,3$.
The relaxation rate of the singlet can become zero, but all other
states attain a minimum relaxation rate of $2\gamma$.

Several other properties of the dynamical map (\ref{eq: 2qbit lindblad})
become evident when we consider the rate of decrease in purity of
an initially pure state; the map ensures that the purity $p(t)=\mathsf{Tr}[\rho^{2}(t)]$
is a monotonically decreasing function of time. A general equation
for the rate of change of purity follows straightforwardly from (\ref{eq: 2qbit lindblad})
as\begin{eqnarray}
\frac{dp}{dt} & = & -3\gamma(1-\xi)p-2\gamma\xi\mathsf{Tr}[S^{2}\rho^{2}]+\gamma\mathsf{Tr}\left[(1+\xi)(\mathbf{S}\rho)\cdot(\mathbf{S}\rho)+(1-\xi)(\mathbf{\Delta}\rho)\cdot(\mathbf{\Delta}\rho)\right].\label{eq:pdot gen}\end{eqnarray}
 For pure initial states $\rho(0)=\left|\psi\right\rangle \left\langle \psi\right|$,
the initial loss in purity occurs at the rate\begin{eqnarray}
\dot{p}(0) & = & -3\gamma(1-\xi)-2\gamma\xi\left\langle \psi\right|S^{2}\left|\psi\right\rangle +\gamma(1+\xi)\left\Vert \left\langle \psi\right|\mathbf{S}\left|\psi\right\rangle \right\Vert ^{2}+\gamma(1-\xi)\left\Vert \left\langle \psi\right|\mathbf{\Delta}\left|\psi\right\rangle \right\Vert ^{2}.\label{eq:purity 0}\end{eqnarray}
 It is easily verified that the separable states of the general form
$\left|a\right\rangle \left|b\right\rangle $ lose initial purity
at the rate $\dot{p}(0)=\gamma(-3+2\left\Vert \left\langle a\right|\mathbf{S}_{1}\left|a\right\rangle \right\Vert ^{2}+2\left\Vert \left\langle b\right|\mathbf{S}_{2}\left|b\right\rangle \right\Vert ^{2})$,
which is independent of $\xi$. Thus the size of $\xi$ has no effect
on the initial decay in purity of unentangled states. Furthermore,
since any state corresponds to {}``spin up'' in some direction,
we have $\left\langle a\right|\mathbf{S}_{1}\left|a\right\rangle \cdot\left\langle a\right|\mathbf{S}_{1}\left|a\right\rangle =\left\langle b\right|\mathbf{S}_{2}\left|b\right\rangle \cdot\left\langle b\right|\mathbf{S}_{2}\left|b\right\rangle =1/4$,
and all separable states are on an equal footing with respect to the
initial rate of loss in purity. 

We now show that this is in fact the minimum rate achievable within
the triplet manifold. An arbitrary state in the triplet manifold has
$\left\langle \psi\right|S^{2}\left|\psi\right\rangle =2$, and $\left\langle \psi\right|\mathbf{\Delta}\left|\psi\right\rangle =0$,
which when substituted in (\ref{eq:purity 0}) yields the initial
rate $\dot{p}(0)=-\gamma(3+\xi-(1+\xi)\left\Vert \left\langle \psi\right|\mathbf{S}\left|\psi\right\rangle \right\Vert ^{2})$
with only one state dependent variable, $\left\Vert \left\langle \psi\right|\mathbf{S}\left|\psi\right\rangle \right\Vert $.
The maximum value of $\left\Vert \left\langle \psi\right|\mathbf{S}\left|\psi\right\rangle \right\Vert $
occurs for a separable $\psi$, which consists of parallel spins.
Therefore, we see that within the triplet manifold, separable states
are the most robust against loss in purity.

On the other hand, the singlet has $|\dot{p}(0)|=3\gamma(1-\xi)$,
which becomes less than $2\gamma$ for $\xi>1/3$. Therefore the singlet
is more robust than the separable-states in this regime. The rate
corresponding to any superposition of singlet and separable states
is greater than the average of the rates of these states. Consequently,
the set of robust states does not change in any continuous manner
with $\xi$, and it instead changes from separable to singlet at $\xi=1/3$.
For $t>0$, the fidelity, $f(t)\equiv\mathsf{Tr}[\rho(t)\rho(0)]$,
offers a characterization for predictability in which $f(t)\approx1$
identifies a highly predicable state and $f(t)\approx0$ a poorly
predicable one. In a study of fidelity, not reported here, we found
that the set of most predicable states makes a transition from separable
to singlet state for $1/3<\xi<1/2$, depending on the time elapsed. 

\begin{figure}

\caption{\label{fig:at sep}The {}``instantaneous rate'' $a(t)$ for (a)
$\xi=0$ and (b) $\xi=0.7$}

\subfigure[]{\includegraphics[%
  scale=0.45]{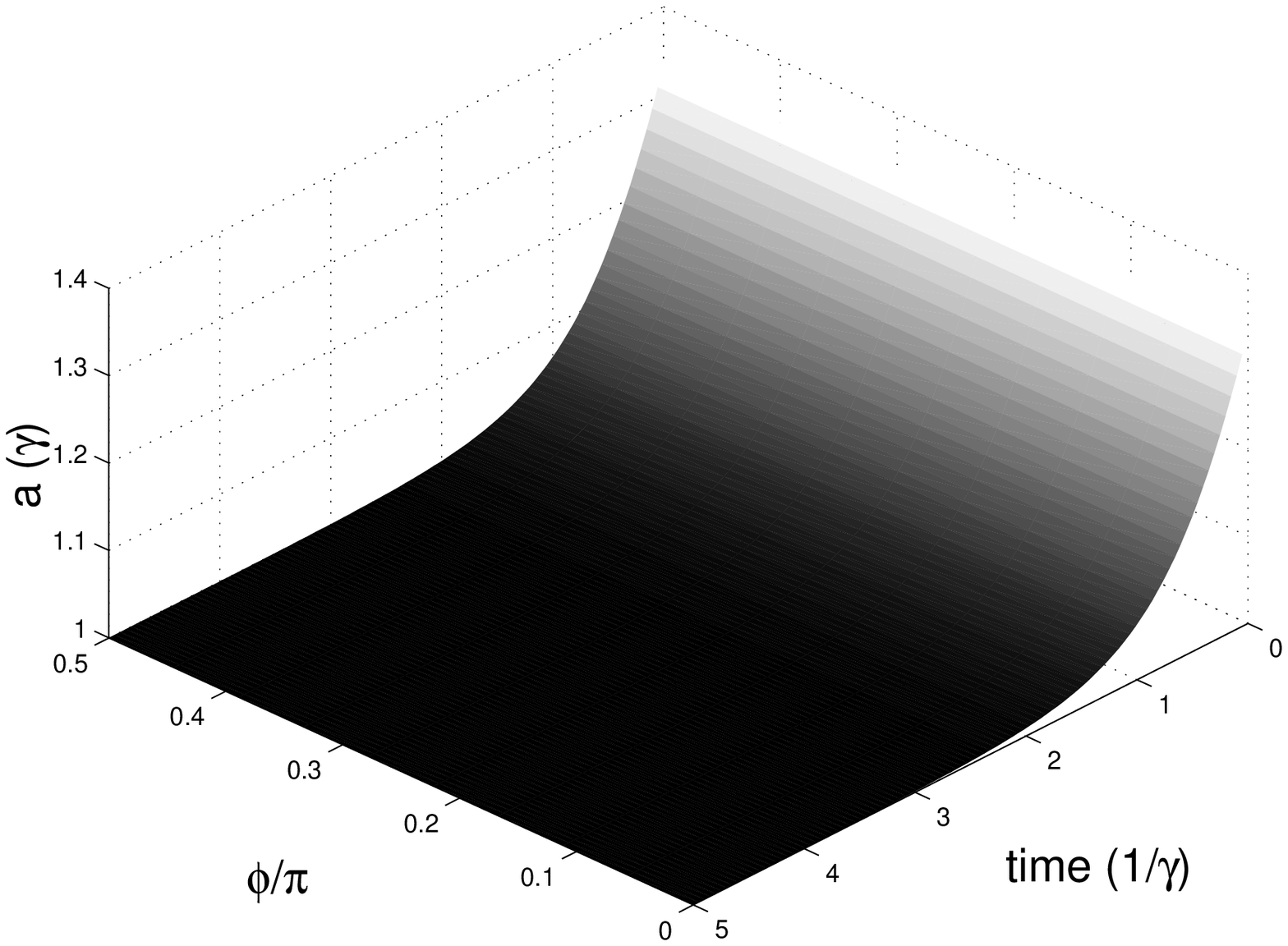}} \subfigure[]{\includegraphics[%
  scale=0.45]{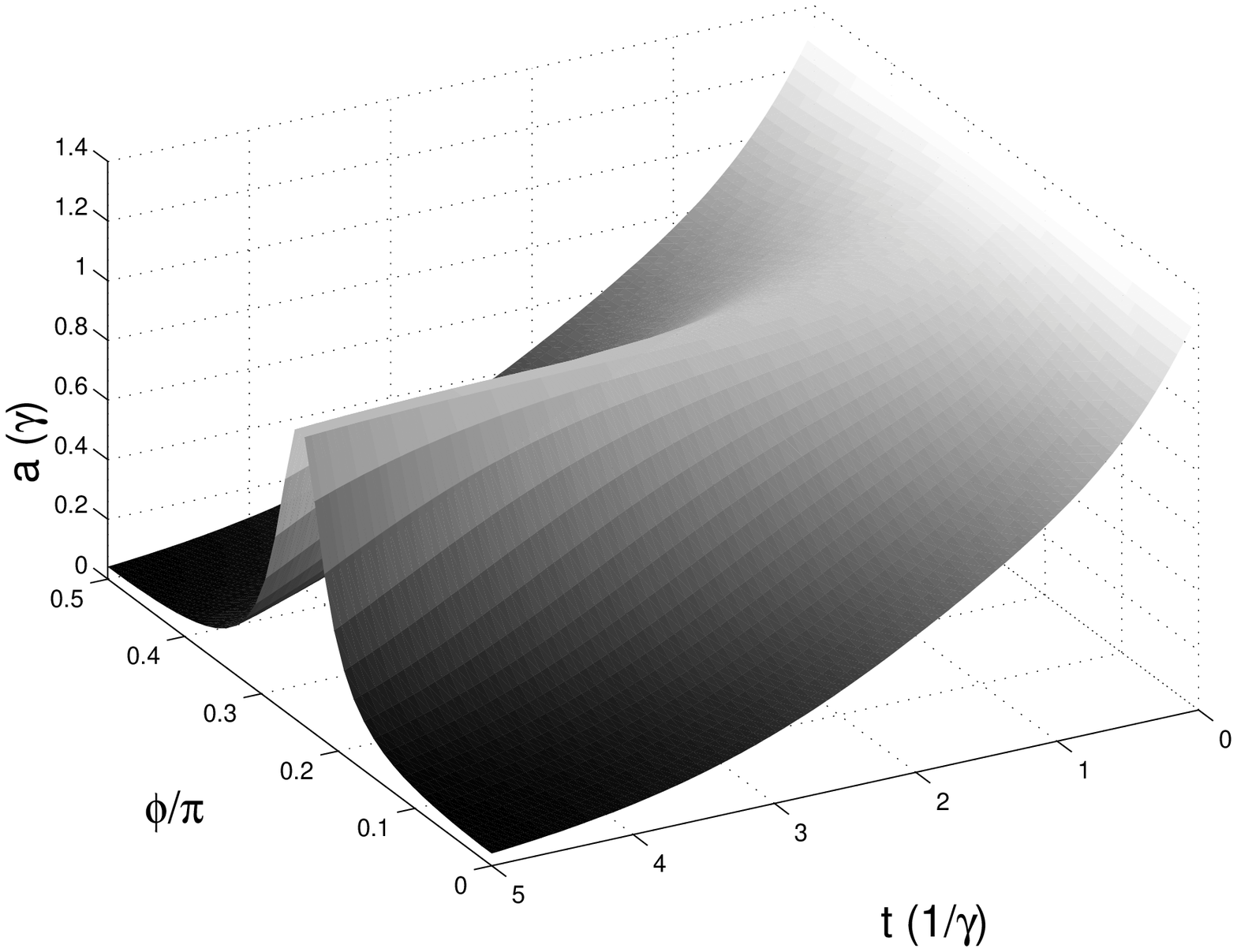}}
\end{figure}

We end this section by discussing the time-dependent behavior of
$\dot{p}(t)$ for a few simple cases. Since for our Hilbert space
of dimension 4, $p(t)$ decreases monotonically to the value $1/4$
for $\xi\neq1$, it is useful to define an {}``instantaneous rate''
for purity\begin{eqnarray*}
a(t) & = & -\frac{\dot{p}}{p-0.25},\end{eqnarray*}
 in terms of which \begin{eqnarray*}
p(t) & = & \frac{1}{4}+\frac{3}{4}\exp\left(-\int_{0}^{t}a(t')dt'\right).\end{eqnarray*}
 The function $a(t)$ highlights differences in $\dot{p}(t)$, and
it is especially useful when different states lead to similar behavior
for $\dot{p}(t)$. We first consider the separable states of the form\begin{eqnarray*}
\left|\psi\right\rangle  & = & \left|0\right\rangle (\cos(\phi)\left|0\right\rangle +\sin(\phi)\left|1\right\rangle )\end{eqnarray*}
 and plot the $\phi$-parameterized rate $a_{\phi}(t)$ as a function
of $\phi$ and time $t$ for different values of $\xi$. Figure (\ref{fig:at sep}a)
shows that all separable states give rise to exactly the same rate
at all times for $\xi=0$. Here we set $\omega=0$ to describe the effects of non-unitary dynamics only.
 The effect of non-zero exchange coupling
becomes evident in Figure (\ref{fig:at sep}b) where we set $\xi=0.7$.
Here the function $a_{\phi}(t)$ decays much more slowly with time
for states $\phi\approx\pi/4$ than for $\phi\in\{0,\pi/2\}$. Thus
while $\xi$ plays no role initially for the separable states, the
more stable states are those with both qubits prepared parallel or
anti-parallel to each other (see discussion after (\ref{eq:purity 0})).
Similarly, the most vulnerable of separable states are those in which
the qubits are eigenstates of spin operators corresponding to orthogonal
Cartesian directions.

The purity of the four Bell states has the following time dependence:\begin{eqnarray*}
p_{i}(t) & = & \frac{1}{4}+\frac{1}{12}e^{-4\gamma(1-\xi)t}+\frac{2}{3}e^{-2\gamma(2+\xi)t},\,\,\,\, i=1,2,3\\
p_{4}(t) & = & \frac{1}{4}+\frac{3}{4}e^{-4\gamma(1-\xi)t}.\end{eqnarray*}
 Thus the three states with total spin $S\neq0$ lose purity for all
values of $\xi$, and do so at two different rates for $\xi\neq1$.
Initially, the decay is dominated by the rate $2\gamma(2+\xi)$, while
at times much longer than the inverse of this rate, the decay approaches
the slower rate $4\gamma(1-\xi)$. The corresponding rate $a(t)$
decreases from $4\gamma(1+\xi/3)$ to $4\gamma(1-\xi)$ as time increases.
In contrast, the rate for the singlet is independent of time.

\section{Conclusion}

In summary, we have derived and studied the master equations for impurity
bound electrons (qubits) scattering a bath of conduction electrons
in a semiconductor. We show that the distribution of bath electrons
remains Boltzmann at all temperatures, due to the temperature dependent
filling of the conduction band. Thus our analysis, based on this result,
is in contrast with other studies on decoherence of qubits in an electron
gas obeying a Fermi-Dirac distribution \cite{RKKY,flips}. The master
equations are obtained in the Lindblad form for a single qubit and
a system of two mutually non-interacting qubits. In the former, the
Lindblad operators are found to be the spin operators for the qubit.
In the latter, these are replaced by the sum and difference of the
spin operators of each member of the system. The Bloch equations derived
for a single qubit show that decoherence occurs at the same rate as
relaxation in the $S^{z}$ basis. This departs from the conventional
inequality $2T_{2}\leq T_{1}$ because the two levels in the system
are degenerate, and therefore relaxation and dephasing processes are
both elastic. The inverse relaxation times are equal to $\gamma$,
which is proportional to the product of thermal flux of the bath electrons
and an effective cross section that depends on the exchange coefficient
$J$ and the Bohr radius of the qubit. Calculations show that $\gamma^{-1}$
is on the order of seconds for temperatures below 10 K, and decreases
rapidly to below a microsecond above 70 K. Thus the conduction electron
bath is dominant in causing decoherence compared to other sources
\cite{kane,rogerio_sarma,x_sarma,loss_vincenzo,koiller_sarma} for
temperatures above 70K. 

The same parameter $\gamma$ also sets the timescale of decoherence
and relaxation in a two qubit system. However an additional parameter,
$\xi(R_{T})$, representing the indirect exchange coupling of the
two members of the system, affects the rates profoundly. The function
$\xi(R_{T})$ decreases monotonically from one to zero with $R_{T}$,
the ratio of inter-qubit distance to thermal deBroglie wavelength
of conduction electrons. The dissipative part of the dynamical map
is dominated by the total spin operator when $\xi\approx1$. As a
result the singlet state, with zero total spin, becomes the most robust
state in this limit. For small $\xi$, however, the separable states
in which both members of the system are in an eigenstate of the same
Cartesian component of the spin operator exhibit the slowest rate
of loss in purity. Pairs of Bell states are found to dephase independently
of each other, and their dephasing rate never exceeds the population
transfer rate. The unitary RKKY interaction arises naturally between
the two qubits in our master equation. The frequency shift associated
with this interaction is found to be proportional to the free electron
susceptibility of the bath, in agreement with past studies of RKKY
interaction between two spins mediated by a gas of free electrons
\cite{kim95,larsen}. While these studies found the interaction to
decay exponentially as a function of inter-qubit distance for a Fermi-Dirac
distribution, we find a Gaussian decay for a Boltzmann distributed
electron gas. The results in the two qubit system can be understood
qualitatively in terms of the coherence length of the bath electrons
and the initial entanglement between the qubits. Electrons with large
thermal deBroglie wavelength tend to scatter as if the two qubits
were acting as a single entity, while those with small wavelength
scatter off each qubit independently of the other. Similarly, qubits
prepared in pure separable states lose purity independently of $\xi$,
while the sensitivity to $\xi$ is much greater for entangled initial
states.

The generalization to include the effects of an external magnetic
field is straightforward and will restore the inequality $2T_{2}\leq T_{1}$
in addition to the precession of qubits in the field. However, interactions
among the qubits demand a more involved calculation, because unless
they are much larger or much smaller than the system-bath interaction,
they render the secular approximation invalid. This approximation
is central to most derivations of a coarse-grained, Markovian master
equation. 

\begin{acknowledgments}
This work was supported by the Natural Sciences and Engineering Research
Council of Canada (NSERC). We thank Eugene Sherman for helpful conversations.
\end{acknowledgments}
\appendix

\section{Derivation of master equation}

Here we derive the master equation (\ref{eq: sec2 rho1}) of section
2. The density operator, denoted $W(t)$, for the joint qubit-electron
system evolves unitarily via the transformation $W(t)=U(t)W(0)U^{\dagger}(t)$,
where the unitary operator $U=1+iT$ and $T$ satisfies the equation\begin{eqnarray}
T & = & -\frac{1}{\hbar}\int_{0}^{t}H_{I}(t')dt'-\frac{i}{\hbar}\int_{0}^{t}H_{I}(t')T(t')dt'.\label{eq:app T}\end{eqnarray}
 By unitarity of $U(t)$ it follows that $i(T-T^{\dagger})=-T^{\dagger}T$,
and the evolution of $W(t)$ can then be written in terms of $T$
as\begin{eqnarray}
W(t)-W(0) & = & \frac{i}{2}[T+T^{\dagger},W(0)]+TW(0)T^{\dagger}-\frac{1}{2}T^{\dagger}TW(0)-\frac{1}{2}W(0)T^{\dagger}T.\label{eq:app W T}\end{eqnarray}
 We consider only the product initial state, $W(0)=\rho_{0}\otimes R_{B}$,
where $R_{B}$ is the thermal density matrix of the bath. We first
show that under the second order Born approximation and appropriate
coarse-graining of $t$, (\ref{eq:app W T}) yields the following
equation for the system density matrix:

\begin{eqnarray}
\rho(t)-\rho_{0} & = & -i\sum_{ijkl}tN_{ijkl}[\eta_{ij}\eta_{kl},\rho_{0}]+\sum_{ijkl}tM_{ijkl}(2\eta_{ij}\rho_{0}\eta_{kl}-\eta_{kl}\eta_{ij}\rho_{0}-\rho_{0}\eta_{kl}\eta_{ij}).\label{eq:app reference eq}\end{eqnarray}
 The tensors $N_{ijkl}$ and $M_{ijkl}$ are independent of time,
and $\eta_{ij}$ are system operators defined by $\eta_{ij}=\left|i\right\rangle \left\langle j\right|$. 

We derive each sum in (\ref{eq:app reference eq}) from the corresponding
term in (\ref{eq:app W T}), correct to second order. Since the expansion
of $T$ starts at the first order in $H_{I}$, the dissipative term
becomes second order automatically. Application of the first order
expansion of $T$ then yields the second sum in (\ref{eq:app reference eq})
with the tensor $M_{ijkl}$ given by the expression: \begin{eqnarray}
tM_{ijkl} & = & \frac{1}{2\hbar^{2}}\int_{0}^{t}dt'\int_{0}^{t}dt''\sum_{\alpha,\alpha'}\sum_{\mathbf{p},\mathbf{p}'}e^{i(\omega_{p}-\omega_{p'})(t''-t')}\left\langle \alpha\mathbf{p};i\right|V\left|\alpha'\mathbf{p}';j\right\rangle \left\langle \alpha'\mathbf{p}';k\right|V\left|\alpha\mathbf{p};l\right\rangle \mathcal{N}_{\mathbf{p}},\label{eq:app M tensor 1}\end{eqnarray}
 where $\mathcal{N}_{\mathbf{p}}$ is the number of particles in the
momentum state $\mathbf{p}$. We now evaluate the integrals and sums
in the limit of a large crystal. Let $\Omega$ represent the crystal
volume, and within this volume, let $n(\mathbf{p})$ be the density
of conduction electrons in the phase space. Then $\mathcal{N}_{\mathbf{p}}=\Omega\Delta\mathbf{p}n(\mathbf{p})$,
where $\Delta\mathbf{p}=(2\pi\hbar)^{3}/\Omega$ is the volume in
$\mathbf{p}$-space associated with a momentum state. Introducing
the phase space volume for $\mathbf{p}'$ in a similar manner and
letting $\Omega$ be large we find that\begin{eqnarray*}
\sum_{\mathbf{p}}\mathcal{N}_{\mathbf{p}}\sum_{\mathbf{p}'} & \rightarrow & \int d\mathbf{p}\Omega n(\mathbf{p})\int d\mathbf{p}'\frac{\Omega}{(2\pi\hbar)^{3}}.\end{eqnarray*}

Next we introduce the rescaled operators $\tilde{V}_{ij}=\Omega(2\pi\hbar)^{-3}\left\langle i\right|V\left|j\right\rangle $
and substitute them in (\ref{eq:app M tensor 1}). Making a change
of variables in $t',t''$ we find that

\begin{eqnarray*}
tM_{ijkl} & = & 4\pi^{3}\hbar\sum_{\alpha\alpha'}\int_{0}^{t}d\tau'\int_{-L(t;\tau')}^{+L(t;\tau')}d\tau\int d\mathbf{p}d\mathbf{p}'e^{i(\omega_{p}-\omega_{p'})\tau}n(\mathbf{p})\left\langle \alpha\mathbf{p}\right|\tilde{V}_{ij}\left|\alpha'\mathbf{p}'\right\rangle \left\langle \alpha'\mathbf{p}'\right|\tilde{V}_{kl}\left|\alpha\mathbf{p}\right\rangle ,\end{eqnarray*}
 where $L(t;\tau')=2\tau'$ for $\tau'\leq t/2$ and $L(t;\tau')=2(\tau-\tau')$
for $\tau\leq t/2$. The integral over $\tau$ is essentially a Fourier
transform of a $p-$dependent function and it is expected to drop
quickly for thermal distributions at sufficiently high temperatures
\cite{carmichael}. Therefore, we switch to a coarse-grained timescale
which captures the evolution of the qubit but not the behavior of
the bath correlations. In this limit, $L(t;\tau')$ may be considered
infinite for essentially all $\tau'$ without introducing error in
the integral over $\tau$. The result is\begin{eqnarray}
tM_{ijkl} & = & t8\pi^{4}\hbar\sum_{\alpha\alpha'}\int d\mathbf{p}d\mathbf{p}'n(\mathbf{p})\delta(\frac{p'^{2}-p^{2}}{2m\hbar})\left\langle \alpha\mathbf{p}\right|\tilde{V}_{ij}\left|\alpha'\mathbf{p}'\right\rangle \left\langle \alpha'\mathbf{p}'\right|\tilde{V}_{kl}\left|\alpha\mathbf{p}\right\rangle .\label{eq:app tM}\end{eqnarray}
 The factor $t$ appears from the integral over $\tau'$ due to the
absence of system dynamics; in the presence of system dynamics, $\tau'$
is coarse-grained further to be insensitive to the energy difference
between the system levels. We now integrate over $\mathbf{p}'$ making
use of the formula $n(\mathbf{p})d\mathbf{p}=4\pi p^{2}n(p)dp(d\mathbf{\hat{\mathbf{u}}}/4\pi)$,
where $n(p)$ is the occupancy of energy state $p^{2}/2m$ divided
by phase space cell volume, and $d\hat{\mathbf{u}}$ is an element
of the solid angle centered about $\hat{\mathbf{u}}$. After substituting
these definitions we find that

\begin{eqnarray}
M_{ijkl} & = & 2(2\pi)^{6}(m\hbar)^{2}\int_{0}^{\infty}dp\frac{p}{m}p^{2}n(p)\sum_{\alpha\alpha'}\int\frac{d\mathbf{\hat{\mathbf{u}}}}{4\pi}\frac{d\mathbf{\hat{\mathbf{u}}}'}{4\pi}\left\langle \alpha p\mathbf{\hat{\mathbf{u}}}\right|\tilde{V}_{ij}\left|\alpha'p\mathbf{\hat{\mathbf{u}}}'\right\rangle \left\langle \alpha'p\mathbf{\hat{\mathbf{u}}}'\right|\tilde{V}_{kl}\left|\alpha p\mathbf{\hat{\mathbf{u}}}\right\rangle .\label{eq:app Mtensor}\end{eqnarray}
 In a similar treatment \cite{sipe} it was shown that a more accurate
calculation yields a result obtained by replacing the first order
scattering amplitudes in this formula with their exact counterparts.
We expect the same to hold here, but since the exchange interaction
is often introduced with parameters assumed appropriate for only a
lowest order Born calculation, we do not pursue this issue here. We
finalize this formula by substituting the Boltzmann distribution for
$n(p)$,\begin{eqnarray*}
n(p) & = & n_{c}(2\pi mk_{B}T)^{-\frac{3}{2}}e^{-p^{2}/2mk_{B}T},\end{eqnarray*}
 where we remind the reader that $n_{c}$ is the total number of conduction
electrons present at temperature $T$. It is convenient to introduce
the dimensionless variable $x=(2mk_{B}T)^{-1}p^{2}$ in terms of which
($v=4(2\pi)^{9/2}n_{c}(m\hbar)^{2}\sqrt{k_{B}T/m}$)\begin{eqnarray}
M_{ijkl} & = & v\int_{0}^{\infty}dxxe^{-x}\int\frac{d\mathbf{\hat{\mathbf{u}}}}{4\pi}\frac{d\mathbf{\hat{\mathbf{u}}}'}{4\pi}\sum_{\alpha\alpha'}\left\langle \alpha,p\mathbf{\hat{\mathbf{u}}}\right|\tilde{V}_{ij}\left|\alpha',p\mathbf{\hat{\mathbf{u}}}'\right\rangle \left\langle \alpha',p\mathbf{\hat{\mathbf{u}}}'\right|\tilde{V}_{kl}\left|\alpha,p\mathbf{\hat{\mathbf{u}}}\right\rangle .\label{eq:app M final}\end{eqnarray}
 Our next task is to expand $i\mathsf{Tr}_{B}(T+T^{\dagger})/2$ and
obtain a time-independent expression for $N_{ijkl}$. As the bath
distribution does not depend on spin, the first order term vanishes
by the zero trace property of Pauli matrices. The second order term
then yields,\begin{eqnarray}
\frac{i}{2}\mathsf{Tr}_{B}[T+T^{\dagger},W_{0}] & = & \frac{-1}{2\hbar^{2}}\int_{0}^{t}dt'\int_{0}^{t'}d\tau\mathsf{Tr}_{B}[[H_{I}(t'),H_{I}(t'-\tau)],\rho_{0}\otimes R_{B}].\label{eq:Hcorr2}\end{eqnarray}
 The commutator $[H_{I}(t'),H_{I}(t'-\tau)]$ takes the following
form where the summation is done over all $\mathbf{p}$ and $\alpha$:

\begin{eqnarray*}
[H_{I}(t'),H_{I}(t'-\tau)] & = & \eta_{ij}\eta_{kl}\sum e^{i(\omega_{p''}-\omega_{p})t'}\left(e^{i(\omega_{p'}-\omega_{p})\tau}-e^{-i(\omega_{p'}-\omega_{p''})\tau}\right)\\
 &  & \times\left|\alpha\mathbf{p}\right\rangle \left\langle \alpha\mathbf{p}\right|V_{ij}\left|\alpha'\mathbf{p}'\right\rangle \left\langle \alpha'\mathbf{p}'\right|V_{kl}\left|\alpha''\mathbf{p}''\right\rangle \left\langle \alpha''\mathbf{p}''\right|.\end{eqnarray*}
 We convert the summations over momentum to integrals and substitute
the result in (\ref{eq:Hcorr2}). Taking the trace we find

\begin{eqnarray}
\frac{i}{2}\mathsf{Tr}_{B}[T+T^{\dagger},W_{0}] & = & [\eta_{ij}\eta_{kl},\rho_{0}]t(-4\pi^{3}\hbar)\sum_{\alpha,\alpha'}\int d\mathbf{p}d\mathbf{p}'\nonumber \\
 &  & \int_{0}^{\infty}d\tau2i\sin[(\omega_{p'}-\omega_{p})\tau]\left\langle \alpha\mathbf{p}\right|\tilde{V}_{ij}\left|\alpha'\mathbf{p}'\right\rangle \left\langle \alpha'\mathbf{p}'\right|\tilde{V}_{kl}\left|\alpha\mathbf{p}\right\rangle n(\mathbf{p}).\label{eq:Hcorr3}\end{eqnarray}
 The integral over $\tau$ yields $2i\mathcal{P}(\omega_{p}-\omega_{p'})^{-1}$,
where $\mathcal{P}$ denotes the principal value. Comparing (\ref{eq:Hcorr3})
with (\ref{eq:app reference eq}), we get\begin{eqnarray}
N_{ijkl} & = & 4\pi^{3}\hbar\int d\mathbf{p}d\mathbf{p}'\frac{2\mathcal{P}}{\omega_{p'}-\omega_{p}}n(\mathbf{p})\sum_{\alpha,\alpha'}\left\langle \alpha\mathbf{p}\right|\tilde{V}_{ij}\left|\alpha'\mathbf{p}'\right\rangle \left\langle \alpha'\mathbf{p}'\right|\tilde{V}_{kl}\left|\alpha\mathbf{p}\right\rangle ].\label{eq:app N 1}\end{eqnarray}
 Substitution of the Boltzmann distribution and re-expression in terms
of the dimensionless variable defined above yields the following expression:

\begin{eqnarray}
N_{ijkl} & = & \frac{v}{\pi}\int dxdy\sqrt{xy}e^{-x}\frac{\mathcal{P}}{y-x}\int\frac{d\mathbf{\hat{\mathbf{u}}}}{4\pi}\frac{d\mathbf{\hat{\mathbf{u}}}'}{4\pi}\sum_{\alpha,\alpha'}\left\langle \alpha\mathbf{p}\right|\tilde{V}_{ij}\left|\alpha'\mathbf{p}'\right\rangle \left\langle \alpha'\mathbf{p}'\right|\tilde{V}_{kl}\left|\alpha\mathbf{p}\right\rangle .\label{eq:app N final}\end{eqnarray}

Having shown the validity of (\ref{eq:app reference eq}), we can
obtain a coarse-grained differential equation for $\rho(t)$ by iterating
this equation after replacing $\rho(t)-\rho(0)$ by $\rho(t+\delta t)-\rho(t)$.
Thus

\begin{eqnarray}
\frac{d\rho}{dt} & = & -i\sum_{ijkl}N_{ijkl}[\eta_{ij}\eta_{kl},\rho]+\sum_{ijkl}M_{ijkl}(2\eta_{kl}\rho\eta_{ij}-\eta_{ij}\eta_{kl}\rho-\rho\eta_{ij}\eta_{kl}).\label{eq:app drhodt}\end{eqnarray}

When the expressions (\ref{eq:V 1}) and (\ref{eq:V 2}) are substituted
for $V$, the tensor $M$ becomes $\gamma\Lambda$, where $\gamma$
is a constant defined in (\ref{eq:sec2 gamma}) and $\Lambda$ is
given by (\ref{eq:lambda}) for the single qubit and by (\ref{eq:lambda q2})
for the two qubit system. The commutator associated with the tensor
$N$ vanishes for the single qubit. In the two qubit system, the tensor
$N=\omega_{RKKY}\Gamma$, where $\omega_{RKKY}$ is defined in (\ref{eq:sec2 omega rkky}),
while $\Gamma$ is given by (\ref{eq:gamma 2}). 

We now outline the calculation to obtain the RKKY splitting in terms
of the susceptibility of the bath. We first observe that (\ref{eq:app N 1})
can be written as \begin{eqnarray*}
N_{ijkl} & = & 4\pi^{3}\hbar\int d\mathbf{p}d\mathbf{p}'2\Re\left(\frac{1}{\omega_{p'}-\omega_{p}+i\epsilon}\right)n(\mathbf{p})\sum_{\alpha,\alpha'}\left\langle \alpha\mathbf{p}\right|\tilde{V}_{ij}\left|\alpha'\mathbf{p}'\right\rangle \left\langle \alpha'\mathbf{p}'\right|\tilde{V}_{kl}\left|\alpha\mathbf{p}\right\rangle ],\end{eqnarray*}
where $\epsilon\rightarrow0$ at the end of the calculation. Substituting
(\ref{eq:V 2}) in the above expression, keeping only the cross-terms,
and changing the variables of integration to $\mathbf{k}=\mathbf{p}/\hbar$,
we find that\begin{eqnarray*}
N_{ijkl} & = & \Gamma_{ijkl}4\pi^{3}\hbar^{-2}(Jr_{0}^{3})^{2}\int\frac{d\mathbf{k}}{(2\pi)^{3}}\frac{d\mathbf{k}'}{(2\pi)^{3}}n(\mathbf{k})e^{i(\mathbf{k}'-\mathbf{k})\cdot\mathbf{R}}2\Re\left(\frac{1}{\omega_{k'}-\omega_{k}+i\epsilon}\right)+c.c.\end{eqnarray*}
Here $\mathbf{k},\mathbf{q}$ are wave-vectors, $n(\mathbf{k})d\mathbf{k}$
now represents the density of electrons with wave-vector within $d\mathbf{k}$
of $\mathbf{k}$, and $\Gamma_{ijkl}$ represents the summation over
spin indices $\alpha,\alpha'$. We now manipulate the sum by first
writing it as a sum of two identical copies of itself and then interchanging
$\mathbf{k},\mathbf{k}'$ in one of the integrals. Then doing the
transformation $\mathbf{k},\mathbf{k}'\rightarrow-\mathbf{k},-\mathbf{k}'$
in that integral, we find that\begin{eqnarray*}
N_{ijkl} & = & \Gamma_{ijkl}4\pi^{3}\hbar^{-2}(Jr_{0}^{3})^{2}\left[\int\frac{d\mathbf{k}}{(2\pi)^{3}}\frac{d\mathbf{k}'}{(2\pi)^{3}}e^{i(\mathbf{k}'-\mathbf{k})\cdot\mathbf{R}}n(\mathbf{k})\Re\left(\frac{1}{\omega_{k'}-\omega_{k}+i\epsilon}\right)\right.\\
 &  & \left.-\int\frac{d\mathbf{k}}{(2\pi)^{3}}\frac{d\mathbf{k}'}{(2\pi)^{3}}e^{i(\mathbf{k}'-\mathbf{k})\cdot\mathbf{R}}n(\mathbf{k})\Re\left(\frac{1}{\omega_{k'}-\omega_{k}+i\epsilon}\right)+c.c\right].\end{eqnarray*}
We do not change the sign of $i\epsilon$ in the second integral since
the real part is unaffected by it. An expression equivalent to the
one above is\begin{eqnarray*}
N_{ijkl} & = & -\Gamma_{ijkl}4\pi^{3}\hbar^{-1}(Jr_{0}^{3})^{2}\int\frac{d\mathbf{q}}{(2\pi)^{3}}e^{i\mathbf{q}\cdot\mathbf{R}}\int\frac{d\mathbf{k}}{(2\pi)^{3}}\frac{[n(\mathbf{k}+\frac{1}{2}\mathbf{q})-n(\mathbf{k}-\frac{1}{2}\mathbf{q})]}{\hbar^{2}\mathbf{k}\cdot\mathbf{q}/m+i\epsilon}+c.c.\end{eqnarray*}
The integral over $\mathbf{k}$ in the limit $\epsilon=0$ is the
static Lindhard function \cite{a/m}, which defines the Fourier transform,
$\chi(\mathbf{q})$, of the static susceptibility $\chi(\mathbf{r})$
\cite{kim95,kim99}. Since $\chi(\mathbf{r})=\chi(-\mathbf{r})$,\begin{eqnarray*}
N_{ijkl} & = & -(2\pi)^{3}\hbar^{-1}(Jr_{0}^{3})^{2}\chi(\mathbf{R})\Gamma_{ijkl}.\end{eqnarray*}

\bibliographystyle{apsrev}
\bibliography{bibliography}

\end{document}